\documentclass[a4paper,11pt]{article}
\usepackage{pos}
\usepackage{natbib}

\title{Probing the Extragalactic Background Light with the MAGIC telescopes}

\author*[a]{R. Grau}
\onbehalf{for the MAGIC collaboration}
\author[a]{A. Moralejo}

\affiliation[a]{Institut de Física d’Altes Energies (IFAE), The Barcelona Institute of Science and Technology (BIST), E-08193 Bellaterra (Barcelona), Spain}

\emailAdd{rgrau@ifae.es}

\abstract{The Extragalactic Background Light (EBL) is the accumulated light produced throughout the universe history, spanning the UV, optical, and IR spectral ranges and mostly originating from stars, directly or re-processed by dust. However, measuring the EBL total intensity (beyond the contribution of resolved discrete sources) is challenging due to its faintness compared to foreground diffuse light like zodiacal light. A possible technique exploits the Very High Energy (VHE) photons coming from sources at cosmological distances. VHE photons can interact with the EBL and produce electron-positron pairs, a process that leaves an imprint in the observed gamma-ray spectrum. Determining the EBL with this method requires assumptions on the intrinsic spectrum of the source, which can affect the robustness of EBL constraints. In this contribution, through the use of Monte Carlo simulations, and archival data of the MAGIC telescopes, we have studied the impact that the assumptions so far adopted in the literature have in the estimates of the EBL density, and how the use of more generic ones would modify the results. These studies can impact our understanding of the evolution of the Universe, gamma-ray propagation, and large-scale structure formation.}

\FullConference{XVIII International Conference on Topics in Astroparticle and Underground Physics (TAUP2023)\\
 28.08\_01.09.2023\\
University of Vienna\\}

\begin{document}
\maketitle

\section{Introduction}
The Extragalactic Background Light (EBL) is all the light produced since the beginning of the universe, covering a range of wavelengths from ultraviolet (UV) to optical and infrared (IR). Most of this light comes from stars, either emitted directly (cosmic optical background, COB) or re-radiated after being absorbed by dust (cosmic infrared background, CIB).\\
There are different methods for probing the EBL. Direct measurements are difficult, as there are much brighter local foregrounds (like zodiacal light) in the same wavelength range. Recent studies \cite{Lauer2022} have used the space telescope LORRI, abroad the New Horizons spacecraft at 51.3 au from the Sun to reduce the uncertainty of the background flux levels. They isolated the COB and estimated a flux of 16.37 $\pm$ 1.47 nW m$^{-2}$ s$^{-1}$, at a pivot wavelength of 0.608 $\mu$m, which is larger than (and in tension with) the results obtained with other methods. Another approach uses the number of galaxies obtained with the combination of wide deep surveys and predictions based on galaxy formation and evolution models to get the integrated flux. The downside of this method is that it is not sensitive to hypothetical truly diffuse and/or unknown components of the EBL. We are using the gamma-ray-based method, where the photons coming from very high energy (VHE) sources interact with EBL photons and create electron-positron pairs. Therefore when looking at VHE sources at cosmological distances, we can see this effect as an energy-dependent absorption imprint on their gamma-ray spectra. We can use this to probe all the EBL, independently of its origin, but we have to make assumptions on the intrinsic gamma-ray spectra at the source. The intrinsic gamma-ray spectra represent how we would observe the source from Earth if there was no EBL, essentially the redshifted gamma-ray spectra of the source.\\
Previous results, like the one from \cite{magic2019}, select concave functions to fit the intrinsic spectra of the sources and then do a profile likelihood of the EBL density scale factor ($\alpha$), relative to the one in a given EBL model. 
The formula
\begin{equation}
    \frac{dF_ {obs}}{dE} = \frac{dF_{int}}{dE} e^{-\alpha\tau(E,z) }
\end{equation}
represents how $\alpha$ accounts for EBL absorption, where E is the energy of arrival at Earth (red-shifted), $\frac{dF_{int}}{dE}$ is the source's red-shifted intrinsic spectrum, and $\tau(E,z)$ is the EBL optical depth in the model.\\
Those results typically use Wilks' Theorem \cite{Wilks} in order to get the alpha constraints from the profile likelihood of alpha, but it is not guaranteed that the conditions of applicability of the theorem are fulfilled. The P-values of the best fits are often very low, hinting at the possible presence of hidden systematic errors (e.g. in the instrument response functions of the telescope), or other  deviations of reality from the assumptions (in the EBL model, or the parametrization of the gamma spectrum).
Another problem with the results obtained with this method could be the selection of the fit function as the result obtained can change depending on which one is selected.\\
Our goal is to make Toy Monte-Carlo simulations in order to check the coverages of the results. These coverages will allow us to check the validity of Wilks' theorem for our study and compute the uncertainties of our results more accurately if they deviate significantly from the targeted $68\%$. Additionally, we aim to use less strong assumptions on the intrinsic spectral shape of the source, using a generic concave function instead of using simple parametrizations with 4 or fewer free parameters.

\section{Toy Monte-Carlo Simulation}
In order to check the validity of Wilks' theorem and compute the uncertainties of our results, we have made a Toy Monte-Carlo code that simulates the observation with MAGIC of different Poisson realizations of the same spectra, modeled by a function. Every realization is then analyzed with a Poissonian likelihood maximization, in the same way real data is treated. The likelihood has one term for each bin in reconstructed energy (i), of the form:
\begin{equation*}
    L_{i}(ebl, \theta) = Poisson(g'_{i}(ebl, \theta) +b_{i}; N_{on,i})\,\cdot Poisson(b_{i}/\beta; N_{off,i})\cdot Gauss(g'_{i};g_{i},\Delta g_{i})
    \label{eq:likelihood}
\end{equation*}
where $N_{on, i}$ and $N_{off, i}$ denote the recorded events in reconstructed energy bins (i = 1, . . . , $N_{bins}$). $N_{on}$ corresponds to events around the source, while $N_{off}$ represents background events in three control regions. $g_i$ is the Poisson parameter (mean number) of gammas in the ON-source region for bin $i$, $b_{i}$ is the Poisson parameter of the background in the On-source region and is treated as a nuisance parameter. The factor $\beta$ is the ratio of ON to OFF exposure which in this case is $\beta = 1/3$. $g_{i}'$ is a nuisance parameter accounting for uncertainty in the instrument response function, while $\Delta g_i$ represents the range of values associated with that uncertainty.\footnote{For more details on the likelihood see Appendix A in \cite{magic2019}}\\
In that way, we obtain multiple Likelihood profiles of the alpha parameter for the different realizations, from $\alpha = 0$ to 2 in steps of 0.05. Then we look at the minimum -2LogLikelihood value and the range of $\alpha$ around it where the -2LogLikelihood does not increase more than 1 ($\Delta LogL <= 1$). To get the $1\sigma$ coverage, we check the percentage of realizations (profile likelihoods) that have the true (simulated) value of $\alpha$ ($\alpha = 1$) inside that region.
If Wilks' theorem can be applied this coverage should be $\sim68\%$, $\sim95\%$ for $2\sigma$ (for $2\sigma$ the region we would look at would be $\Delta LogL <= 2^2$), and so on.\\
The P-values obtained with the real data were much smaller than the ones obtained with the simulation, hinting at the possible presence of hidden systematics. We decided to add to the Toy MC as Gaussian systematics in the effective area, independent in each energy bin, but with the same standard deviation (in relative terms). This is a simple way of simulating them, but there are many other.\\

\begin{figure}[h!]
    \centering
    \begin{minipage}[b]{0.52\textwidth}
        \includegraphics[width = \linewidth]{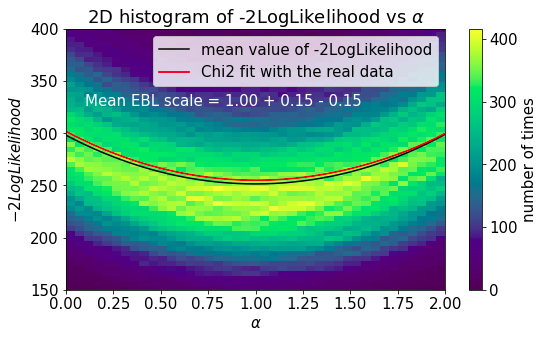}
        \caption{Profile likelihood of 10k realizations of the toy Monte-Carlo simulation for 15 observations of Mrk421 and Dominguez 2011 EBL model. The Gaussian systematics in the effective area added to the simulation are $2.25\%$. Ndoff = 221.}
        \label{fig:sim}
    \end{minipage}
    \hfill
    \begin{minipage}[b]{0.45\textwidth}
        \includegraphics[width = \linewidth]{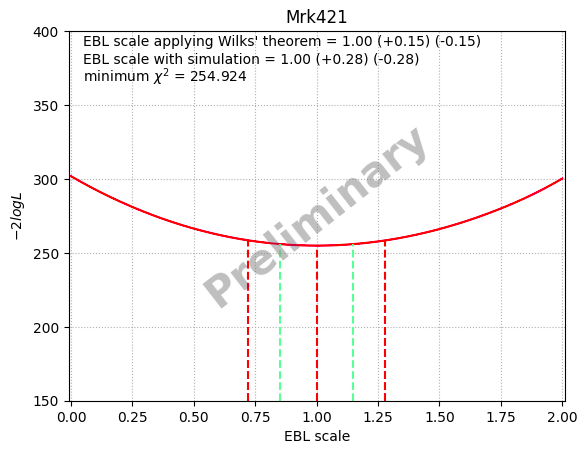}
        \caption{Profile likelihood of the 15 observations of Mrk421 with Dominguez 2011 EBL model. The red dotted lines are the uncertainty computed with the simulation and the green ones with Wilks' theorem.}
        \label{fig:unc}
    \end{minipage}
\end{figure}

We used the Toy MC code to generate 10k realizations of the 15 datasets of Mrk421 emulating the ones of \cite{magic2019} (using the best-fit function of that paper as the intrinsic spectra of each dataset) with the EBL model from \cite{Dominguez}. We also computed their profile likelihood and obtained \autoref{fig:sim}. The coverage obtained with $\Delta LogL <= 1$ is $40.25\%$, lower than the one we would get with Wilks' theorem ($\sim68\%$) and therefore we see that it cannot be applied here. In order to compute the uncertainty of the $\alpha$ value correctly for the real data we will get the $\Delta LogL$ for which a $68.27\%$ of the realizations have the real value of $\alpha$ inside of that region. With that result, we can do the profile likelihood with the real data and compare the two uncertainties obtained, getting \autoref{fig:unc}. We can clearly see how the uncertainty in this case is an $87\%$ larger than the one computed with Wilks' theorem.

\section{Generic concave function}
BLLac objects are not expected to have inflection points in their VHE spectra (after the inverse Compton peak) while the EBL transmissivity $e^{-\tau(E)}$ has an inflection point at about 1 TeV. With the aim of reducing the number of assumptions on the intrinsic spectral shape of the sources analyzed, we decided to use a generic concave function to fit 1ES1011+496 in February 2014, see \cite{magic2019}, which is the individual spectrum of that paper with better lower constraint.\\
The function we decided to use is a Multiply Broken Power Law (MBPWL), a Power Law (PWL) defined by parts where the photon index changes in points called nodes or knots. In order to impose concavity the photon index can only increase. We select the number of knots and the position of the first and last one and then the other knots are logarithmically spaced between the first and last.\\

\begin{figure}[h!]
    \centering
    \includegraphics[width = \linewidth]{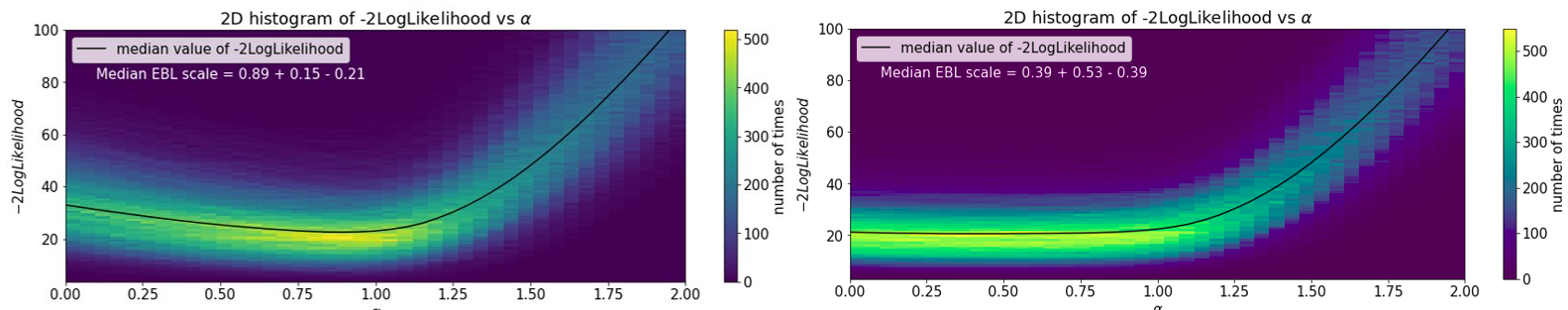}
    \caption{Profile likelihood of 10k realizations of the toy MC simulation of 1ES1011 with Dominguez 2011 EBL model. ON the left we have the fits with the LP (Ndof = 17) and on the right the ones of the MBPWL (Ndof = 16) with 2 nodes.}
    \label{fig:1ES1011}
\end{figure}

We run the toy MC simulating 1ES1011 for 10k realizations and we fitted a Log-Parabola (LP) and a MBPWL with 2 nodes to each one. In \autoref{fig:1ES1011} we can see the profile likelihoods obtained when fitting the LP and the MBPWL for the 10k realizations. This shows how the MBPWL flattens the left part of the curve while the right part is very similar for both functions as both have the concavity constraint and for both functions the best fit becomes a PWL when this limit is reached.

\section{Conclusions}
We made a Toy Monte-Carlo simulation to check if the conditions for applying Wilks' theorem, usually used in previous EBL studies, are fulfilled. After checking the coverages obtained with the Toy MC applying Wilks' theorem, we have seen that there is undercoverage and therefore Wilks' theorem cannot be used in those cases. We used a dataset from a very intense flare with good statistics. Since this case shows undercoverage it is very likely that this effect is present also in other published results using this analytical approach to obtain the uncertainties. We used the Toy MC to calculate more realistic uncertainties for the Mrk421 data that we studied.\\
In order to reduce the number of assumptions done when choosing the source spectral model, to get more robust results, we have studied the use of a generic concave function, the Multiply Broken Power-Law. For the test case of the 2014 1ES1011 flare observed with MAGIC, the use of MBPWL results in weaker lower EBL constraints, while the upper constraint is unchanged. It should be noted that the upper constraint is arguably the most relevant one for the gamma-based EBL determination approach since it sets limits to the possible contributions to EBL beyond those from known galaxy populations.

\acknowledgments
The research leading to these results has received funding from the FSE under the program Ayudas predoctorales of the Ministerio de Ciencia e Innovación  PRE2020-093561

\end{document}